\documentclass[twocolumn]{aastex62}

\usepackage{etoolbox}
\makeatletter 
  \patchcmd{\NAT@citex}
    {\@citea\NAT@hyper@{%
      \NAT@nmfmt{\NAT@nm}%
      \hyper@natlinkbreak{\NAT@aysep\NAT@spacechar}{\@citeb\@extra@b@citeb}%
      \NAT@date}}
    {\@citea\NAT@nmfmt{\NAT@nm}%
    \NAT@aysep\NAT@spacechar\NAT@hyper@{\NAT@date}}{}{}

  \patchcmd{\NAT@citex}
    {\@citea\NAT@hyper@{%
      \NAT@nmfmt{\NAT@nm}%
      \hyper@natlinkbreak{\NAT@spacechar\NAT@@open\if*#1*\else#1\NAT@spacechar\fi}%
        {\@citeb\@extra@b@citeb}%
      \NAT@date}}
    {\@citea\NAT@nmfmt{\NAT@nm}%
    \NAT@spacechar\NAT@@open\if*#1*\else#1\NAT@spacechar\fi\NAT@hyper@{\NAT@date}}
    {}{}
\makeatother
\usepackage{amsmath}
\usepackage{multirow}

\newcommand\stella{\texttt{STELLA}}
\newcommand\sedona{\texttt{Sedona}}
\newcommand\mesa{\texttt{MESA}}

\graphicspath{{./}{figures/}}

\received{May 1, 2020}
\revised{May 28, 2020}
\accepted{May 31, 2020}
\submitjournal{ApJ}

\shorttitle{Type II-P SN RT Modeling Comparison}
\shortauthors{Benny T.-H. Tsang et al.}

\begin{document}

\title{Comparing Moment-Based and Monte Carlo Methods of Radiation Transport Modeling \\
for Type II-Plateau Supernova Light Curves}

\correspondingauthor{Benny T.-H. Tsang}
\email{btsang@kitp.ucsb.edu}

\author{Benny T.-H. Tsang}
\affil{Kavli Institute for Theoretical Physics, University of California, Santa Barbara, CA 93106, USA}

\author{Jared~A. Goldberg}
\affiliation{Department of Physics, University of California, Santa Barbara, CA 93106, USA}

\author{Lars Bildsten}
\affiliation{Kavli Institute for Theoretical Physics, University of California, Santa Barbara, CA 93106, USA}
\affiliation{Department of Physics, University of California, Santa Barbara, CA 93106, USA}

\author{Daniel Kasen}
\affiliation{Departments of Physics and Astronomy, University of California, Berkeley, Berkeley, CA 94720 USA}
\affiliation{Lawrence Berkeley National Laboratory, Berkeley, CA 94720, USA}



\begin{abstract}

Time-dependent electromagnetic signatures from core-collapse supernovae are the
result of detailed transport of the shock-deposited and radioactively-powered radiation
through the stellar ejecta.
Due to the complexity of the underlying radiative processes, considerable 
approximations are made to simplify key aspects of the radiation transport problem.
We present a systematic comparison of the moment-based radiation hydrodynamical 
code \stella\ and the Monte Carlo radiation transport code \sedona\ in the 1D modeling of 
Type II-Plateau supernovae.
Based on explosion models generated from the 
Modules for Experiments in Stellar Astrophysics (\mesa) instrument, 
we find remarkable agreements in the modeled light curves and the ejecta structure thermal evolution,
affirming the fidelity of both radiation transport modeling approaches.
The radiative moments computed directly by the Monte Carlo scheme in \sedona\ 
also verify the accuracy of the moment-based scheme.
We find that the coarse resolutions of the opacity tables and the numerical approximations
in \stella\ have insignificant impact on the resulting bolometric light curves, 
making it an efficient tool for the specific task of optical light curve modeling.

\end{abstract}


\section{Introduction} \label{sec:intro}

Most massive stars with stellar mass above 8\,$M_{\odot}$ end their lives in energetic core-collapse supernova (CCSN) explosions, whose electromagnetic signatures contain information about the progenitor stars and their final moments.
The most common subclass of CCSNe are of Type II-Plateau (II-P) \citep{Li11,Smith11}, explosions of red supergiants (RSG) \citep{WHW02,Heger03,Smartt09,Smartt15,VanDyk12a,VanDyk12b} with prolonged release of shock energy powering the characteristic plateaus in their light curves. 

Current time-domain survey facilities,
e.g., the Zwicky Transient Facility \citep[ZTF,][]{ZTF19} and 
the All-Sky Automated Survey for Supernovae \citep[ASAS-SN,][]{ASASSN17},
as well as the deployment of the Vera C. Rubin Observatory \citep{LSST09},
will significantly increase our capacity to explore SN populations. 
An important goal in these studies will be to reconcile the variety of observed
SN events with the properties of their massive star progenitors and resulting explosions.

In light of the imminent expansion of photometric data, 
the inverse problem of inferring progenitor and explosion properties from observed 
light curves represents a key challenge in achieving 
physical understanding of the full II-P SNe population. 
Developments in 3D simulations have advanced our understanding of RSG explodability and 
observational CCSN signatures \citep{Wongwathanarat15,Janka16,Vartanyan19,Burrows20}, 
but the enormous demands in computing resources limit the exploration of
large suites of multi-dimensional progenitor and explosion models.
Synthetic light curves, on the other hand, can be generated relatively quickly
with one-dimensional (1D) radiation hydrodynamical or 
radiation transport models based on 1D progenitor and explosion models. 

A valuable approach to deriving light curves from RSG progenitors 
is via a combination of radiation hydrodynamical and radiation transport simulations.
In the radiation hydrodynamics step, stellar progenitors of II-P SNe are constructed
with a stellar evolution code, and explosions are simulated 
by either a moving piston or a thermal bomb near the inner boundary
\citep{Woosley95,Woosley07,DH10,DH11}.
The resulting ejecta structures are the initial conditions for the radiation
transport modeling of photometric and spectral features. 
\citet{KW09} adopted this approach to compute the broadband light curves and spectral evolution with the multi-wavelength Monte Carlo radiation transport code \sedona. 
This approach allowed close examination of the dependence of light curve diagnostics
on the progenitor model parameters and the refinement of analytical scaling formulae \citep{Popov93}.

Similar approaches have been applied to build the observable-progenitor 
parameter relation \citep[see, e.g.][]{Dessart13,Morozova16,Sukhbold16}.
More recently, suites of progenitor and explosion models have highlighted the high 
degree of degeneracy between light curve features/photospheric 
velocities and model parameters \citep{DH19,Goldberg19,MB19}. 
While photometric and spectral observables are useful in constraining the
model parameters, they do not offer associations to unique sets of 
progenitor properties after $\approx20$ days. 
Before this time, when the ejecta is still accelerating and the emission comes 
from the outer $<0.1M_\odot$ of material, 
photospheric velocities might distinguish between otherwise degenerate light curves. 
However, such measurements are rare and prone to modifications by the uncertain circumstellar environment (e.g. \citealt{Moriya18}).

Observable signatures from II-P SNe are the results of the complex interplay between 
radiation and the structures, compositions, and opacities of the ejecta. 
The predicted observable characteristics may also be sensitive to the numerical schemes 
used for radiation transport. 
For example, \citet{MB19} and \citet{DH19} both adopted the gray diffusion approximation
during the hydrodynamical evolution step, light curves and spectra in \citet{DH19} 
were produced using a more sophisticated, wavelength-dependent 
non-local thermodynamical equilibrium (non-LTE) scheme.
Discrepancies may also be introduced for other practical reasons, such as the use of 
an opacity floor \citep{Morozova15,MB19}. 

The effects of the choice of radiation transport schemes have been evaluated in 
the context of radiation-driven gas dynamics \citep{KT12,D14,Teyssier15,TM15,ZD17}.
In transient signature predictions, several works have attempted to
verify the accuracy of different numerical radiation transport approaches.
\citet{Kozyreva17} have compared pair-instability SN
light curves produced by different codes.
However, verification works for II-P SNe are very scarce, and they are usually based on a handful of progenitor models and on visual inspections only
\citep{Morozova15,Kozyreva19}.

In this paper, we focus on the II-P SNe and conduct a detailed comparison between
the moment-based radiation hydrodynamics code \stella\ 
and the particle-based Monte Carlo radiation transport code \sedona.
We make direct comparisons of the light curves and ejecta profiles.
We note that the numerical approximations and ray-tracing radiation transport scheme
in \stella\ offer significant speedup over 
\sedona's detailed opacity calculations and Monte Carlo sampling. 
Our goals are to assess and quantify the differences yielded by the two 
fundamentally different radiation transport approaches, to understand the impacts different 
radiation treatments entail, and to offer confidence in the interpretations of 
similar light curve analyses.

In Section \ref{sec:methods}, we discuss the numerical methodology with emphasis 
on the aspects related to radiation transport, and describe our progenitor 
and explosion model suite.
In Section \ref{sec:results}, we present the results of the numerical experiment
and compare the two radiation transport methods.
In Section \ref{sec:summary_discussion}, we summarize our findings and discuss
potential future directions.

\section{Numerical Methods} 
\label{sec:methods}

We use three numerical tools to model the hydrodynamics of the 
massive star progenitor and the radiation transport in the expanding ejecta.
The progenitor evolution and traversal of the shock within the star are 
calculated using \mesa\ \citep{Paxton2011, Paxton2013,Paxton2015,Paxton18,Paxton19}.
The emergent light curves and the ejecta properties are calculated with 
the moment-based, radiation hydrodynamics code \stella\ \citep{Blinnikov98,BS04,BBP05,Blinnikov06}
and the particle-based, Monte Carlo radiation transport code \sedona\
\citep{Kasen06}.

\subsection{Modeling Red Supergiants and their Explosions in \mesa}
\label{sec:RSGs}
The progenitor stellar models were selected from \citet{Paxton18},
the standard suite of models in \citet{Goldberg19},
and models motivated by matching observations described in \citet{Goldberg20}. 

The inlist parameter files for these progenitor models follow those used in \citet{Paxton18}, namely, from the \texttt{test\_suite} case \texttt{example\_make\_pre\_ccsn} with revised values for the progenitor mass, mixing length in the H-rich envelope, initial rotation, core overshooting, and the initial metallicity set to solar ($Z~=~0.02$)

After removing the core, a thermal bomb is injected into the innermost 0.01$M_\odot$ of each model, which heats the star to a specified total final energy $E_{\rm exp}$. The evolution of the shock is then modeled in \mesa\ with the ``Duffell RTI" prescription for mixing from Rayleigh-Taylor Instability \citep{Duffell16,Paxton18}, 
with the modeling terminating at shock breakout.
All models\footnote{Progenitor models not originating from \citet{Paxton18} used \mesa\ revision 10398. Explosions used \mesa\ revision 10925, except for M12.7\_R719\_E0.84\_Ni048, which was carried out in revision 11701.}
remove fallback onto the inner boundary as described in 
\citet{Paxton19} and \citet{Goldberg19}. 
We focus on models at sufficiently high explosion energy that fallback is negligible. 
Radioactive $^{56}$Ni distributions are scaled to match the total 
nickel mass desired.

The \mesa\ models at the moment of shock breakout are handed off to \stella\ as described 
in \citet{Paxton18}. 
Our models have no additional material outside the stellar photosphere. 
The spatial zoning in the \stella\ input files is determined by interpolating across the \mesa\ profiles to match a specified number of zones.
The models are then evolved from near-shock breakout to day 170 
(see \citet{Paxton18} for more details).
We used 400 spatial zones and $N_\textrm{freq} = 40$ frequency bins in \stella, 
as convergence studies \citep{Paxton18} showed sufficient agreement in 
bolometric light curves at this choice of resolution.

\begin{deluxetable*}{lcccccc}[t]
\tablecaption{Summary of the RSG progenitor and explosion models with the corresponding ejecta properties.
The models are named after the concatenation of the ejecta mass, 
progenitor radius at shock breakout, explosion energy, and the 
nickel mass (if any).
The notation follows 
M$[M_\textrm{ej}]$\_R[$R$]\_E[$E_\textrm{Exp}$](\_Ni[$M_{\textrm Ni}$]).
\label{tab:rsg_models}}
\tablecolumns{10}
\tablewidth{0pt}
\tablehead{
\colhead{Model Name} & \colhead{$M_\textrm{ZAMS}$}  & \colhead{$M_\textrm{final}$} & \colhead{$M_\textrm{c,He}$} & \colhead{$M_\textrm{H,tot}$} & \colhead{$X_\textrm{He,env}$} &
\colhead{$t_\textrm{SB}$} \\
\colhead{($M_{\odot}$)($R_{\odot}$)(10$^{51}$\,\textrm{erg})(10$^{-2}$\,$M_{\odot}$)} & \colhead{($M_{\odot}$)} & \colhead{($M_{\odot}$)} & \colhead{($M_{\odot}$)}  & \colhead{($M_{\odot}$)} & \colhead{\,} &
\colhead{(day)}
}
\startdata
\multicolumn{7}{c}{Nickel-Rich Suite} \\
*M9.3\_R433\_E0.5\_Ni1.5 & \multirow{3}{*}{11.8} & \multirow{3}{*}{10.71} & \multirow{3}{*}{3.58} & \multirow{3}{*}{4.65} & \multirow{3}{*}{0.33} & 1.10 \\
*M9.3\_R433\_E1.0\_Ni3.0 &  &  &  &  & & 0.70  \\
*M9.3\_R433\_E2.0\_Ni6.0 &  &  &  &  & & 0.40  \\[0.1cm]
\ddag M12.7\_R719\_E0.84\_Ni4.8 & 15.0 & 14.53 & 5.09 & 6.37 & 0.31 & 1.63  \\ [0.1cm] 
\dag*M16.3\_R608\_E1.0\_Ni4.5 & \multirow{3}{*}{19.0} & \multirow{3}{*}{17.79} & \multirow{3}{*}{5.72} & \multirow{3}{*}{7.53} & \multirow{3}{*}{0.34}  & 1.37  \\
\dag*M16.3\_R608\_E2.0\_Ni3.0 &  &  &  &  & & 0.97  \\
\dag*M16.3\_R608\_E2.0\_Ni7.5 &  &  &  &  & & 0.97 \\
\hline     
\multicolumn{7}{c}{Nickel-Poor Suite} \\
*M9.3\_R433\_E0.5 & \multirow{4}{*}{11.8} & \multirow{4}{*}{10.71} & \multirow{4}{*}{3.58} & \multirow{4}{*}{4.65} & \multirow{4}{*}{0.33} & 1.10 \\ 
*M9.3\_R433\_E0.8 &  &  &  &  & & 0.78  \\ 
*M9.3\_R433\_E1.0 &  &  &  &  & & 0.70  \\ 
*M9.3\_R433\_E2.0 &  &  &  &  & & 0.50  \\ [0.1cm]
\dag*M16.3\_R608\_E1.0 & \multirow{2}{*}{19.0} & \multirow{2}{*}{17.79} & \multirow{2}{*}{5.72} & \multirow{2}{*}{7.53} & \multirow{2}{*}{0.34} & 1.37  \\ 
\dag*M16.3\_R608\_E2.0 &  &  &  &  & & 0.97  \\
\hline
\multicolumn{6}{c}{Exploratory Suite} \\
\ddag M7.9\_R375\_E0.23\_Ni4.3 & 10.0 & 9.41 & 3.15 & 4.06 & 0.33 & 1.13  \\ 
\ddag M16.5\_R533\_E4.6\_Ni13.0 & 19.0 & 18.09 & 6.28 &8.04 & 0.31 & 0.55  \\ 
\dag Stripped\_M4.7\_R379\_E1.0\_Ni3.0 & 17.0 & 6.39 & 6.25 & 0.10 & 0.67 & 0.28\tablenotemark{a}
\enddata
\tablenotetext{}{\textbf{Note:} The data columns are the initial zero-age main sequence stellar mass
$M_\textrm{ZAMS}$, final stellar mass at the time of explosion $M_\textrm{final}$,
pre-explosion helium core mass $M_\textrm{c,He}$, 
total hydrogen mass in the ejecta $M_\textrm{H,tot}$, 
helium mass fraction in the hydrogen-rich envelope $X_\textrm{He,env}$,
and the time from thermal bomb to shock breakout $t_\textrm{SB}$.}
\tablenotetext{\dag}{Progenitor models taken from the \citet{Paxton18} suite.}
\tablenotetext{*}{Models taken from the \citet{Goldberg19} suite.}
\tablenotetext{\dag*}{\ Progenitor models were taken from \citet{Paxton18}, and the explosions 
were performed as part of \citet{Goldberg19}}
\tablenotetext{\ddag}{Models taken from \citep{Goldberg20}}
\tablenotetext{a}{Numerically, shock breakout is defined here as the time when the outgoing shock reaches a small overhead mass coordinate in \mesa\ (0.05\,$M_{\odot}$), which is typically within an hour of the maximum bolometric luminosity in \stella.
Because of the low hydrogen envelope mass in the II-b-like model, the \mesa-to-\stella\  handoff occurred significantly earlier than actual shock breakout. Thus, for this model, $t_\textrm{SB}$ is defined as the time of maximum bolometric luminosity in \stella.}
\end{deluxetable*}

\subsection{Model Selection}

We compiled a characteristic set of models to capture the variations of CCSN progenitor properties.
Models are divided into three suites. 
The \emph{nickel-rich} suite consists of models typical of II-P SNe.
The \emph{nickel-poor} models have no radioactive nickel.
The goal is to compare the radiation transport results in the absence of radioactive power.
We also explore models unlike common II-P SNe in the \emph{exploratory} suite.
In particular, we test models with extreme values of 
$M_{\rm Ni}$ and $E_{\rm exp}$.
The M7.9\_R375\_E0.23\_Ni4.3 model is motivated by SN2009ib, which
had an unusually long plateau \citep{Takats15}; 
the M16.5\_R533\_E4.6\_Ni13 model is informed by the energetic II-P event
SN2017gmr \citep{Andrews19}.
The Stripped\_M4.7\_E1.0\_Ni03 model is not a typical II-P SN --
it is a model with most of its hydrogen stripped to resemble a II-b SN.
We include this model to test the reliability of radiation transport modeling 
when gamma-rays are not fully trapped.

The model names and their key properties of the progenitor and explosion are summarized in Table \ref{tab:rsg_models}.
The time to shock breakout $t_\textrm{SB}$ is included to
indicate the relative expansion times, and the helium fraction in the
envelope $X_\textrm{He,env}$ is provided for reference as helium abundance
can modify the plateau duration and brightness \citep{KW09}.

\subsection{Evolution to Homology and Handoff to \sedona}
\label{sec:homology_handoff}
\sedona\ does not follow ejecta hydrodynamics. Therefore,
models from \stella\ must be handed off as inputs to \sedona,
at a time late enough so that further hydrodynamical evolution 
is insignificant, and early enough to capture meaningful radiation signatures. 

In the course of the evolution with \stella, the outer $\approx$75\% of the 
ejecta mass establishes homology by about 5 days after shock breakout.
In the outer region, assuming homology thereafter introduces $\leq$5\% difference 
in the velocity field from the hydrodynamical calculation.
Beyond $\approx$5 days, the largest deviations from homology appear only near 
the inner boundary of the ejecta, where the reverse shock 
is sometimes still moving inward.

By default, \sedona\ evolves the ejecta assuming homology,
i.e. radius of the $i$th spatial zone evolves as $r_{i}(t_{\rm exp}) = v_{i}\,t_{\rm exp}$,
where $v_{i}$ is the \emph{time-independent} zone velocity and 
$t_{\rm exp}$ is the time since explosion.
However, the conventional $r = v\,t$ homology assumption does not 
strictly apply for Type II-P SNe due to the large progenitor radii.
We modified \sedona\ to incorporate this difference -- by setting 
$r_{i}(t') = r_{i, {\rm handoff}} + v_{i}\,t'$, 
taking into account its initial radius at model handoff, 
and $t'$ is the time since handoff. 
\sedona\ then evolves the ejecta density structure assuming 
that the mass contained in each zone remains constant,
and the zone volume is updated by our revised radius relation.

We need to ensure that the velocity profiles of the ejecta models
are sufficiently time-invariant.
The actual time after the \stella-to-\sedona\ handoff when homology in the
entire ejecta is established depends on the specific models.
We choose day 20 after shock breakout as the \stella-to-\sedona\ handoff time, 
as by then most of the ejecta energy is in kinetic rather than internal energy.
This choice is consistent with the timescale for establishing homology 
found in \citet{Utrobin17} for II-Ps with a different 1D code \texttt{CRAB}.
In all our models, the time to shock breakout $t_\textrm{SB}~<~2$\,days.
Taking $t_\textrm{SB}$ to be the characteristic timescale for the SN expansion,
model handoff at day 20 corresponds to $>10$ expansion times after the explosion. 

At the day-20 handoff, radial profiles from \stella, 
which contain the zone radius, radial velocity, gas density and temperature, 
and the abundances of 15 atomic species,
are copied onto a 1D spherically symmetric, Lagrangian grid in \sedona\ 
without regridding.
In other words, \sedona\ sees an identical ejecta on the same grid as \stella\ 
at the moment of handoff. 
Once handed off, \sedona\ performs Monte Carlo radiation transport and
ejecta evolution fully independent of \stella\ up to day 200 (150) 
for nickel-rich/exploratory (nickel-poor) models.

\subsection{Radiation Transport in \stella\ and \sedona}
\label{sec:RT}

\stella\ solves the time- and frequency-dependent moment equations of 
radiation intensity on a Lagrangian 1D spherical grid.
The system of two moment equations are closed by a variable Eddington factor, 
which is computed by integrating the time-independent transport equation 
for each frequency bin using a ray tracing integration scheme \citep{ZS94,Blinnikov98}.

A low number of logarithmically-spaced frequency bins $N_{\rm freq} = 40$ is used to 
span the wavelength range $1-50,000$\,\AA.
The radiation source terms are coupled to the 
equations of hydrodynamics and solved implicitly \citep{Blinnikov98}.
For thermal radiation, the following sources of opacity are taken into account:
bound-free/photoionization \citep{Verner93}, free-free \citep{GM78}, 
bound-bound/line opacity \citep{Kurucz95,VVF96}, and electron scattering.
Bound-bound opacity is included using the Sobolev approximation under
the line expansion formalism \citep{Karp77,EP93}.
The Sobolev line optical depth of the $i$th bound-bound
transition is given by
\begin{align}
    \tau_{\rm sob, i} &= \frac{\pi e^{2}}{m_\textrm{e} c} f_{i} n_{i} \lambda_{i} \cdot t_{\rm exp}, \label{eqn:tau_sob}
\end{align}
where $f_{i}$ is the oscillator strength of the transition, 
$n_{i}$ is the number density of the lower atomic level of the transition,
$\lambda_{i}$ is the line center rest wavelength, and the
$t_{\rm exp}$ term follows from the velocity gradient of the homologously expanding SN ejecta.
In our models, $t_\textrm{exp}$ is taken to be the time since shock breakout,
i.e., \mesa-to-\texttt{STELLA} handoff.
In this formalism, the large velocity gradient Doppler shifts photons 
into resonance with each lines in a frequency bin exactly once, 
and the lines are treated as non-overlapping.
Line opacity is further assumed to be fully absorptive, 
an approximation justified by non-LTE calculations \citep{Baron96,PintoEastman00}.

For gamma-rays, \stella\ uses the one-group approximation, which effectively 
treats the transport and deposition of gamma-ray energy with a gray absorption opacity
\citep{SSH95}. Furthermore, kinetic energy of the positrons resulting from $^{\rm 56}$Co
decays is assumed to be deposited locally \citep{Blinnikov06}.
The bolometric light curves are computed by summing the radiative fluxes 
over all frequency bins.

In the public version of \stella, opacity tables are constructed on a fixed grid of 
density and temperature, with $\log(\rho/\rm{g\,cm}^{-3})$ spanning from
-18 to -4, and $\log(T/\rm{K})$ from 3.4 to 6.2, 
each with $N_{\rm grid}~=~14$ grid points uniformly separated in the logarithmic space.
To keep the calculations tractable, opacity tables are only computed 
at six times at 
$(t/\rm{days}) \in \{1, 2.5, 6.3, 15.8, 39.8, 100\}$ after shock breakout, 
and on a reduced number of $N_{\rm reduced}=50$ uniformly spaced spatial zones.
In the more recent, private version of \stella\ \citep{Blinnikov06},
the last two approximations have been removed and the opacity is computed
at every time step for every zone, but these new features are not yet publicly available.
Ionization fractions and level populations of the 15 tracked atomic species are 
followed using the LTE approximation.
A simplification is made to include only the six most populated ionization stages.
The pre-computed opacity tables are interpolated in the course of \stella's 
radiation hydrodynamical evolution.

The default $\log(T/\textrm{K})$ resolution of the opacity tables is too
coarse across the hydrogen recombination temperature
($\log(T_{\rm recomb}/\rm{K}) \approx 3.7$) and it
reduces the accuracy of radiation transport in
II-P ejecta as the opacity varies sharply across the recombination front.
We thus modified the public version to allow 
a finer opacity grid with $N_{\rm grid} = 56$ and $N_{\rm reduced}=200$. 
In addition, we appended the stimulated emission correction term,
$1 - \exp{(hc/\lambda_{i} k_{\rm B} T)}$,
to the Sobolev optical depth of the line expansion opacity, Equation (\ref{eqn:tau_sob}).
These improvements lead to smoother ejecta structures and bolometric 
light curves, and they will be available in an upcoming \mesa\ release.
In the following, modeling results from our modified version of \stella\ are 
presented unless otherwise specified.

\begin{figure*}
    \centering
    \includegraphics[width=0.75\textwidth]{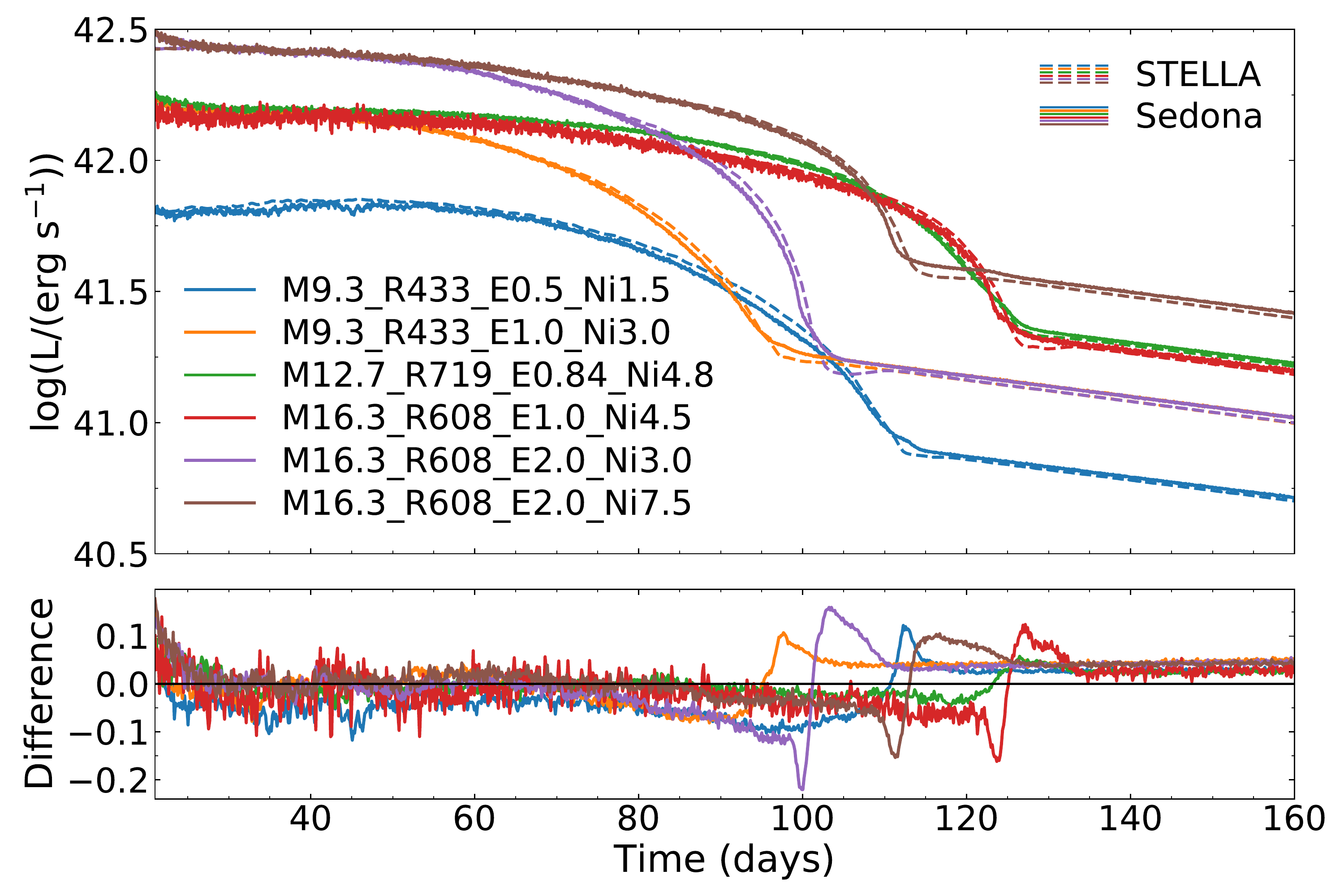}
    \includegraphics[width=0.75\textwidth]{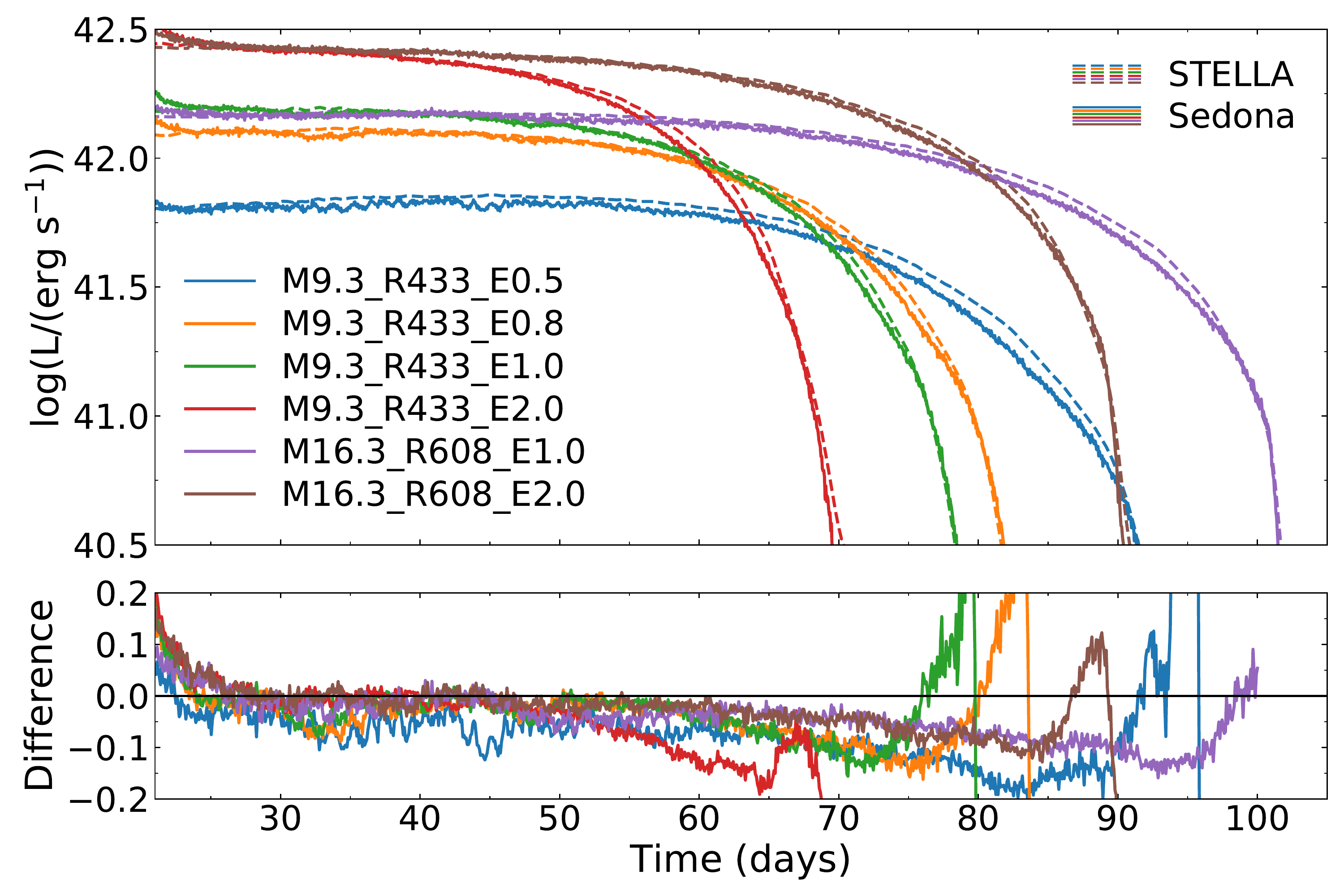}
    \caption{Bolometric light curves from \sedona\ (solid) and \stella\ (dashed)
             for the nickel-rich (top) and nickel-poor (bottom) suites.
        Relative differences between the bolometric luminosities computed by
        \sedona\ and \stella, 
        defined as $(L_\texttt{Sedona} - L_\texttt{STELLA}) / L_\texttt{Sedona}$,
        are shown in the lower panels.
        } \label{fig:lc_comparison}
\end{figure*}

Time-dependent, multi-frequency radiation transport is also performed 
with the Monte Carlo code \sedona\ \citep{Kasen06}.
Monte Carlo methods directly solve the radiation transfer equation 
by discretizing the radiation field into individual photon packets.
Sampling the emission, absorption, and scattering of the
packets through space and time yields a statistical approximation to
the solution of the transfer equation.
Monte Carlo methods are fundamentally different from 
moment-based methods.
We adopt the mixed-frame formalism for the Monte Carlo radiation transport
\citep{MK82} --  
emissivity and opacity are defined in the fluid co-moving frame,
while the quantities are Lorentz-transformed into the lab frame 
where radiation transport is carried out. 
Unlike \stella, \sedona\ is optimized to handle much denser frequency grids.
However, to facilitate direct comparison, we used frequency
limits and resolution that match \stella's default.

The temperature structure is evolved in \sedona\ by considering the expansion,
the radioactive heating from the decay of $^{56}$Ni and $^{56}$Co,
and radiative transport in each zone.
The ionization structures and level populations are determined assuming LTE.
For thermal radiation, the same types of opacity are included, namely, 
bound-free, free-free, bound-bound, and electron scattering.
We note that the bound-free and free-free opacities are computed from a 
different set of analytical formulae than that used by \stella.
Following the common practice, we assume line opacity to be purely absorptive.
A large number of $N_\textrm{init} = 10^{7}$ Monte Carlo photon packets are used to represent the 
initial thermal radiation field at model handoff.
An additional $N_\textrm{nuc} = 10^{5}$ packets are added per radiation transport 
time step to sample the radioactive luminosity in those models containing $^{56}$Ni. 

The Monte Carlo radiation treatment allows \sedona\ to self-consistently 
handle the frequency-dependent transport and thermalization of gamma-rays 
\citep{Kasen06}.
Positrons produced during the nuclear decays are assumed to be 
thermalized instantaneously into photons,
but their energies can be deposited non-locally depending on
the opacity along the photons' trajectories.

To properly set the time step size for the homology evolution in \sedona, 
we consider the relevant timescales of radiation transport.
In the lab frame where \sedona's radiation transport is performed, 
the radiative flux comprises both the diffusion and advection contributions. 
In spherical symmetry, the transformation to $\mathcal{O}(v/c)$ is
\citep{MihalasMihalas99}, 
\begin{align}
    F_{\rm r} &= F_{\rm r,0} + v(r) \cdot (E_{\rm r,0} + P_{\rm r,0}), \label{eqn:LT_flux} \\
              &= F_{\rm diff,0} + F_{\rm adv,0},
\end{align}
where $E_{\rm r,0}$, $F_{\rm r,0}$, and $P_{\rm r,0}$ denote the radiation energy
density, flux, and pressure measured in the co-moving fluid frame, 
$v(r)$ is the radial velocity, 
and $F_{\rm r}$ without the subscript `$0$' denote the lab frame radiative flux.
The two terms on the right hand side of Equation (\ref{eqn:LT_flux}) can be
respectively identified as the diffusive and advective flux.
The characteristic timescales for radiative diffusion and advection 
across the local shell thickness of the spherical grid, $\Delta r$, are
\begin{align}
    t_{\rm diff} &= E_{\rm r,0} \Delta r/ F_{\rm diff,0}, \\
    t_{\rm adv} &= E_{\rm r,0} \Delta r / F_{\rm adv,0}.
\end{align}
In the optically thick regime, these expressions simplify by invoking the
Fick's law of radiation transfer and assuming isotropy ($P_{\rm r,0} = E_{\rm r,0} / 3$),
\begin{align}
    t_{\rm diff} &\approx E_{\rm r,0} \Delta r \left[ \frac{c}{3} \frac{d E_{\rm r,0}}{d\tau} \right]^{-1}  \approx \frac{3}{c} \tau(r) \Delta r, \label{eqn:t_diff}\\
    t_{\rm adv} &\approx \frac{3}{4} \frac{\Delta r}{v(r)} \label{eqn:t_adv},
\end{align}
where $\tau(r)$ is the optical depth to infinity from radius $r$.
By equating $t_\textrm{diff}$ and $t_\textrm{adv}$ in
Equation (\ref{eqn:t_diff}) and (\ref{eqn:t_adv}),
we obtain a characteristic optical depth 
$\tau_\textrm{char} = c/4v$
below which radiation is no longer locally trapped by diffusion 
and radiation transport is expected to modify the ejecta's thermal
structures.

In SNe, the dynamics of the ejecta are dominated by the expansion.
For the \sedona\ models, we therefore chose a constant time step size 
of 0.05\,day, which is $\lesssim$0.1\,$t_{\rm adv}$
throughout the entire duration of the simulations.
Light curves are generated by tallying the Monte Carlo photon packets that
escape the ejecta through the outer boundary.

\begin{deluxetable*}{lcccccc}[t]
\tablecaption{Comparison of light curve diagnostics from the \stella\ and \sedona\ modeling.
The plateau duration $t_\textrm{p}$ is computed using the same procedure 
as in \citet{Goldberg19}.
\label{tab:lc_results}}
\tablecolumns{7}
\tablewidth{0pt}
\tablehead{
\colhead{\,} & \multicolumn{3}{c}{{$L_{\rm 50}$/10$^{42}$\,{\rm erg\,s}$^{-1}$}} & \multicolumn{3}{c}{{$t_{\rm p}$/{\rm days}}} \\
\colhead{Model} & \colhead{\stella}  & \colhead{\sedona} & \colhead{\% Difference} & \colhead{\stella} & \colhead{\sedona} & \colhead{\% Difference}
}
\startdata
\multicolumn{7}{c}{Nickel-Rich Suite} \\
M9.3\_R433\_E0.5\_Ni1.5 & 0.70 & 0.67 & -4.3 & 104.2 & 106.7 & 2.3 \\
M9.3\_R433\_E1.0\_Ni3.0 & 1.371 & 1.374 & 0.2 & 89.16 & 87.75 & -1.6 \\
M9.3\_R433\_E2.0\_Ni6.0 & 2.17 & 2.25 & 3.9 & 81.9 & 80.2 & -2.1 \\ [0.1cm]
M12.7\_R719\_E0.84\_Ni4.8 & 1.54 & 1.53 & -0.3 & 116.6 & 119.5 & 2.4 \\ [0.1cm]
M16.3\_R608\_E1.0\_Ni4.5 & 1.46 & 1.49 & 2.2 & 120.8 & 119.8 & -0.8 \\
M16.3\_R608\_E2.0\_Ni3.0 & 2.43 & 2.39 & -1.7 & 97.4 & 96.6 & -0.8 \\
M16.3\_R608\_E2.0\_Ni7.5 & 2.45 & 2.45 & $<10^{-3}$ & 108.5 & 107.9 & -0.5 \\
\multicolumn{7}{c}{Nickel-Poor Suite} \\
M9.3\_R433\_E0.5 & 0.70 & 0.66 & -6.2 & 93.8 & 95.7 & 2.0 \\
M9.3\_R433\_E0.8  & 1.18 & 1.15 & -2.1 & 84.0 & 83.8 & -0.2 \\
M9.3\_R433\_E1.0  & 1.37 & 1.34 & -2.1 & 83.6 & 80.0 & -3.1 \\
M9.3\_R433\_E2.0  & 1.99 & 1.93 & -3.4 & 70.2 & 70.5 & 0.5 \\ [0.1cm]
M16.3\_R608\_E1.0  & 1.48 & 1.39 & -5.9 & 105.0 & 102.6 & -2.3 \\
M16.3\_R608\_E2.0 & 2.45 & 2.42 & -1.3 & 90.0 & 91.5 & 1.7 \\
\hline
\enddata
\end{deluxetable*} 

\begin{figure}
    \includegraphics[width=\columnwidth]{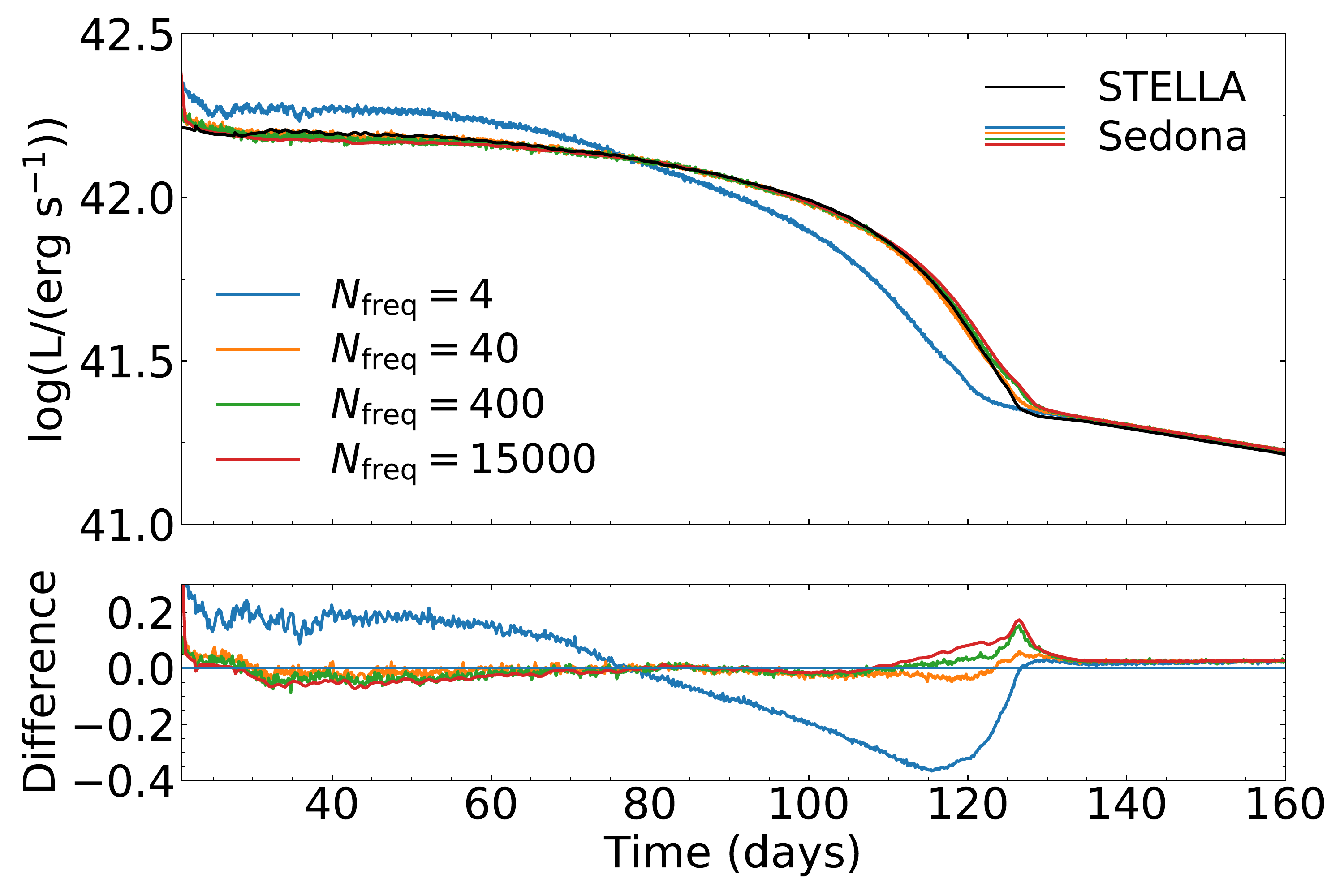}
    \caption{Bolometric light curves of the M12.7\_R719\_E0.84\_Ni4.8 model
    using different numbers of frequency bins. The wavelength range for radiation transport is held fixed at 1$-$50000\,\AA\ for this
    numerical experiment. 
    In both codes, the frequency grids are uniform in $\log$(frequency) space.
    The light curves converge once $N_\textrm{freq} \ge 40$, the default value
    in \stella. 
    Relative differences between the light curves from \sedona\ and \stella, 
    defined as $(L_\texttt{Sedona} - L_\texttt{STELLA}) / L_\texttt{Sedona}$,
    are shown in the bottom panel.} \label{fig:lc_Nfreq_comparison}
\end{figure}

\begin{figure*}
    \includegraphics[width=\columnwidth]{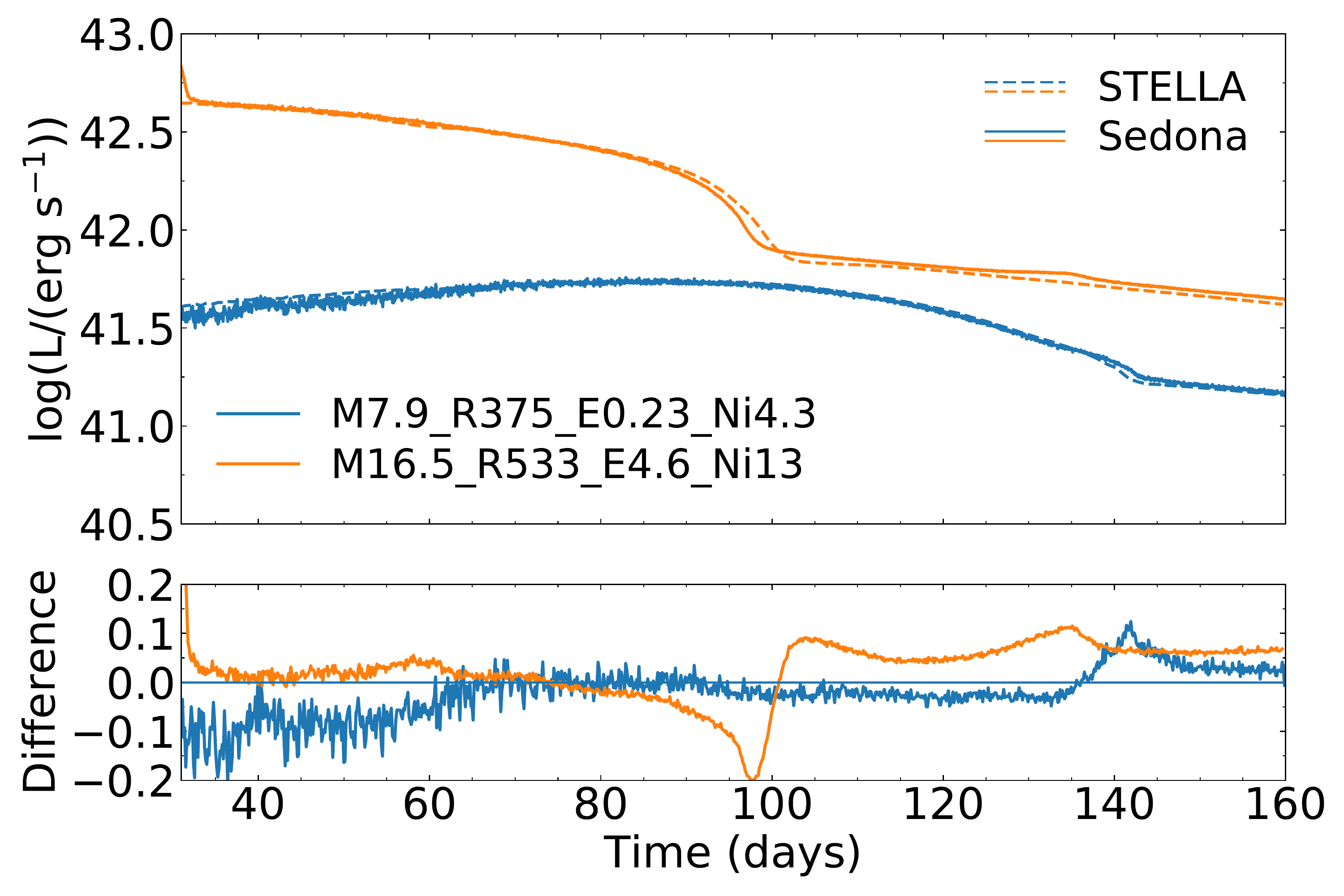}
    \includegraphics[width=\columnwidth]{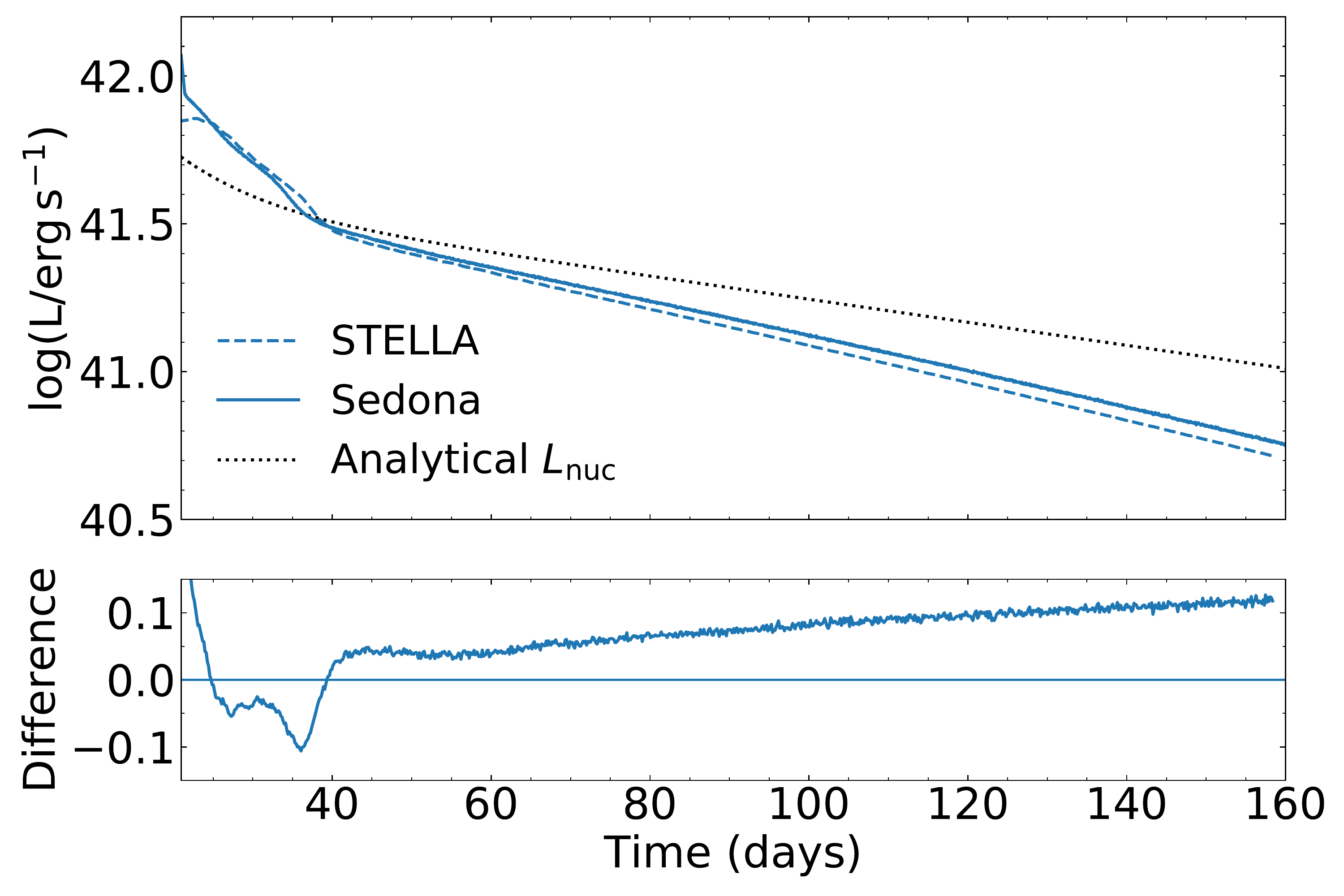}
    \caption{Bolometric light curves from the exploratory suite. 
    The II-P models M7.9\_R375\_E0.23\_Ni4.3 and M16.5\_R533\_E4.6\_Ni13
    with extreme $M_\textrm{ej}/E_\textrm{Exp}$ are shown in the left panel.
    The II-b-like model Stripped\_M4.7\_E1.0\_Ni03 model is shown on the right. 
    The black dotted line on the right panel shows the total nuclear decay energy production rate 
    integrated over the entire ejecta.
    The relative differences between the \sedona\ and \stella\ light curves, 
    defined as $(L_\texttt{Sedona} - L_\texttt{STELLA}) / L_\texttt{Sedona}$,
    are shown in the bottom panels.
        } \label{fig:explore_lcs}
\end{figure*}

\begin{figure*}
\includegraphics[width=\columnwidth]{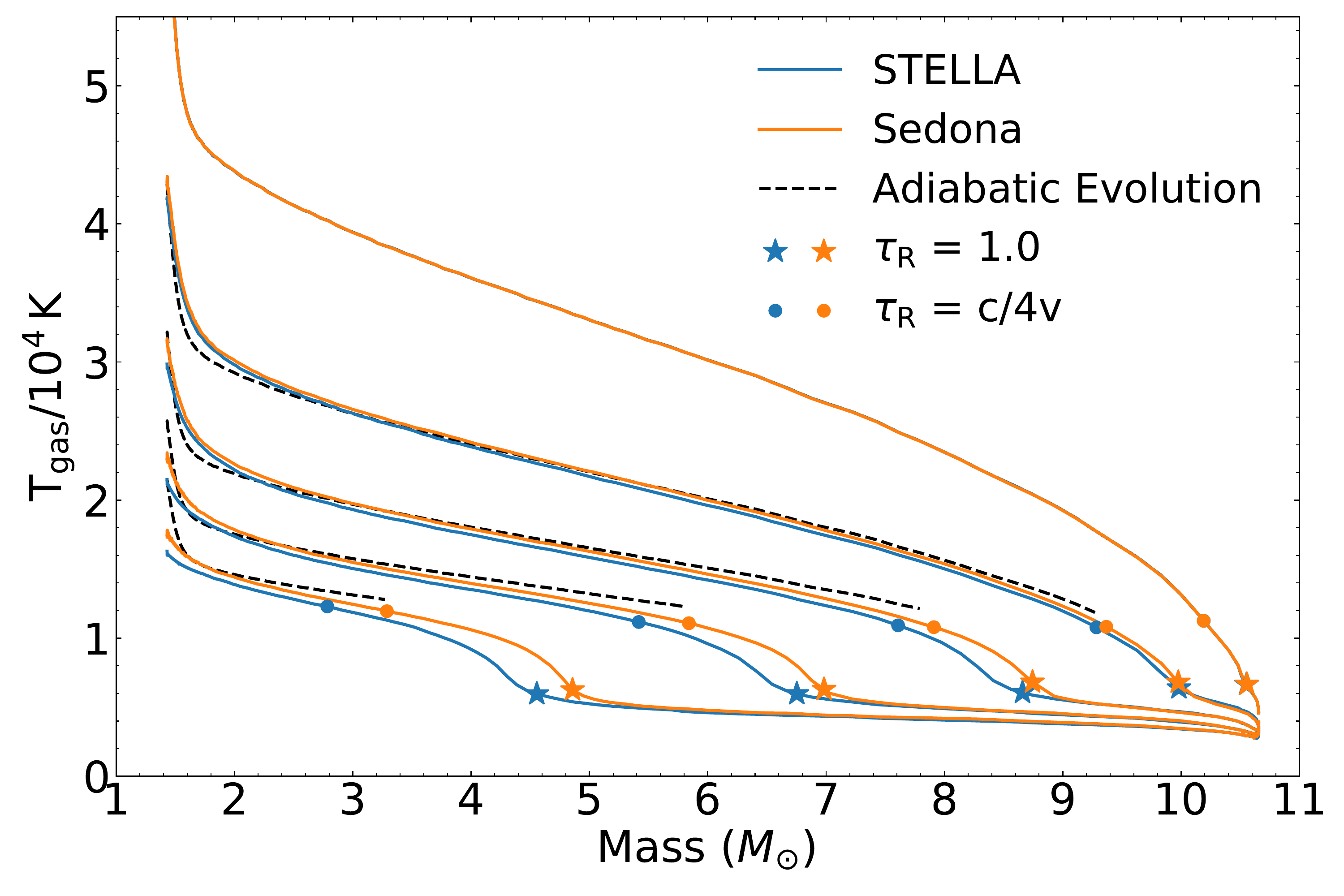}
\includegraphics[width=\columnwidth]{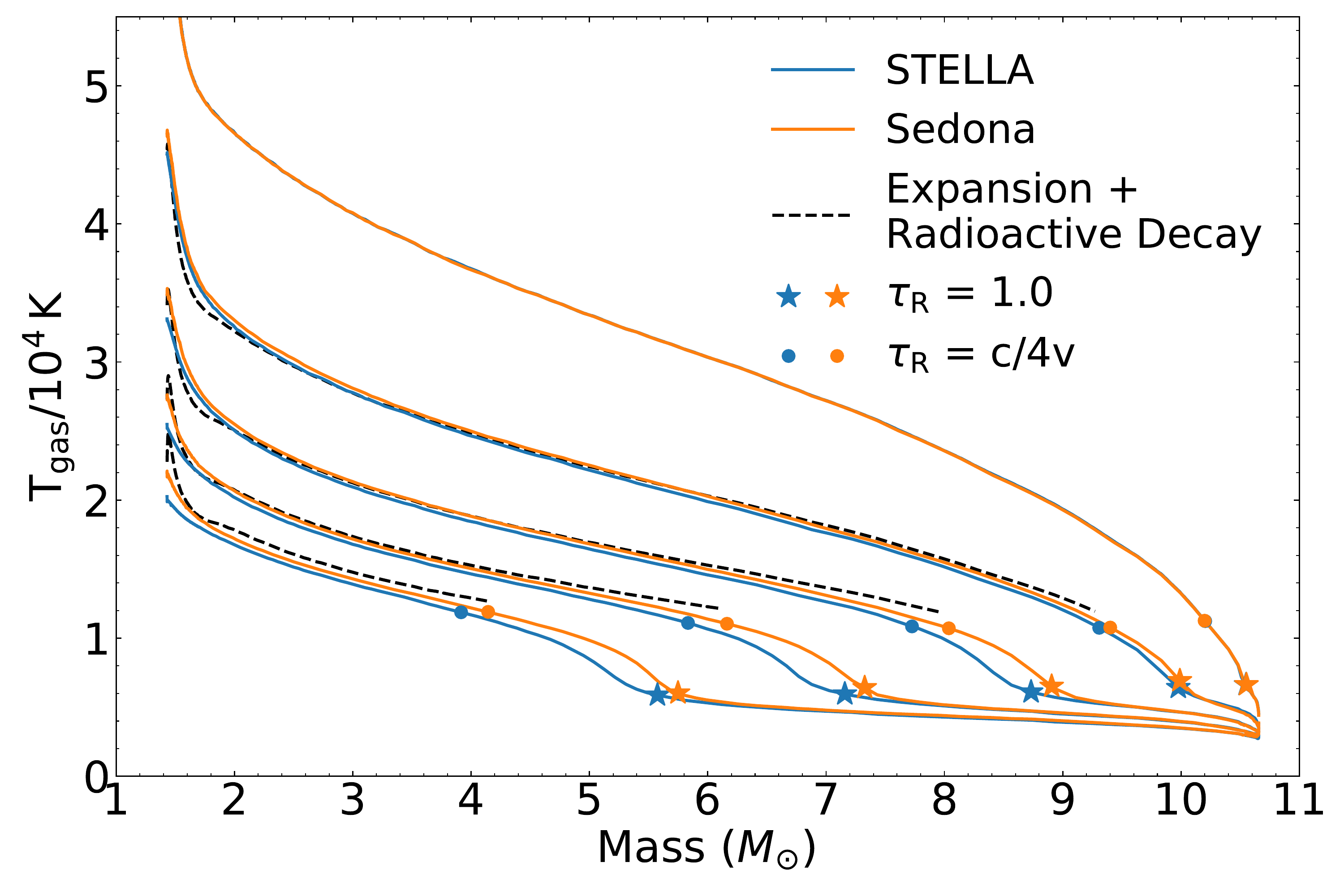}
\caption{Gas temperature evolution of the M9.3\_R433\_E1.0(\_Ni3.0) model is displayed 
        on the left (right) panel.
         The lines correspond to, from top to bottom, profiles at
         day 20, 30, 40, 50, and 60.
         The stars mark the locations of the photosphere. 
         The circular points denote where $\tau_\textrm{R} = c/4v$.
         The black dashed lines are the analytical scaling relations -- 
         on the left panel we use adiabatic evolution, 
         $T_{\rm gas}(t) = T_{\rm gas}({\rm day\,20}) (t / 20\,{\rm days})^{-1}$;
         on the right the total local deposition of radioactive decay energy is assumed \citep[Equation (13)]{Nadyozhin94}).
         Both analytical relations are truncated at $\tau_\textrm{R} = c/4v$.
         } \label{fig:temp_profile_comparison}
\end{figure*}

\section{Results} \label{sec:results}

\subsection{Bolometric Light Curves}

The top panels of Figure \ref{fig:lc_comparison} compare the light curves from 
the representative models in the nickel-rich and nickel-poor suites,
as computed by \sedona\ (solid) and \stella\ (dashed).
The lower panels below the light curves show the relative difference between the two.
The moment-based radiation transport scheme of \stella\ matches well 
the Monte Carlo approach of \sedona\ following handoff at day 20.
During the plateau phase, the bolometric luminosities between the 
two methods agree within 5\%.
The systematic bias of \stella\ models towards higher luminosity during the plateau
is from its hydrodynamical evolution. 
Following handoff, a small fraction of internal energy is still being converted 
into kinetic energy.
At day 100, the residual acceleration results in a $\approx$3\% increase in radius
in the majority of the ejecta compared to a constant-velocity evolution,
giving rise to a $\approx$10\% reduction in density as compared to \sedona\ 
(see also Section \ref{sec:ej_structures}).
The lower overall density in the \stella\ models reduces the ejecta optical depth,
allowing more radiation to escape and enhancing the luminosity.

To quantify the difference between the light curves, 
we compare the two common diagnostics, day-50 luminosity $L_\textrm{50}$ and the plateau duration $t_\textrm{p}$,
in Table \ref{tab:lc_results}. 
We use the same procedure as in \citet[Equation (9)]{Goldberg19} 
to obtain $t_\textrm{p}$.
Both $L_\textrm{50}$ and $t_\textrm{p}$ agree to within 2--6\%, 
offering confidence in the reliability of both radiation transport approaches. 

At the end of the plateaus in the nickel-rich models, 
slight dips are observed on the \stella\ light curves, causing differences of about 10--20\%.
Similar dips on the light curves were also observed in previously published \stella\ models
\citep{Paxton18,Kozyreva19,Goldberg19},
but the underlying cause was not discussed.
At the times of the dips, we observe luminosity declines in the outer ejecta 
where the optical depths to the surface are below $10^{-3}$.
At this time, the opacity in the entire ejecta is dominated by 
neutral atomic lines, which by default are assumed to be purely absorptive. 
Since the dips never manifested in \sedona\ models, we believe that they 
are produced by artificial absorption in extremely optically thin regions in \stella.
We confirmed that such dips do not occur when the line opacity is switched
to purely scattering. 

\sedona's Monte Carlo approach to gamma-ray transport enables an independent 
verification of \stella's gray approximation for the gamma-ray radiation.
For the nickel-rich models, 
the total deposition rate of decay energy integrated over 
the entire ejecta follows the total decay luminosity to within $\leq$2\%.
At day 150, the column densities of electrons
$N_\textrm{e} = \int n_\textrm{e}(r)dr$~$\ge$\,10$^{26}$\,cm$^{-2}$,
equivalent to a Compton scattering optical depth of 
$\tau_{sc} = N_\textrm{e} \sigma_\textrm{KN} \sim 100$ for 1\,MeV gamma-ray photons,
where $\sigma_\textrm{KN}$ is the Klein-Nishina-corrected scattering cross section.
Therefore, local deposition of decay energy within the ejecta is a
reasonable approximation for the overall energetics and light curve modeling.
However, in agreement with \citet{Wilk19}, we observe in the \sedona\ models that 
there are significant spatial and temporal variations of the gamma-ray deposition rate 
throughout the ejecta.

In the radioactive decay phase, \sedona\ persistently gives a luminosity 3--5\%
higher than that of \stella. 
This is due to two different aspects of the numerical treatment of radioactive decay.
First, the numerical constants of radioactive energy released per decay used in 
\stella\ and \sedona\ differ by 1--2\%.
The constants adopted in \texttt{STELLA} and \texttt{Sedona} are taken from 
\citet{Nadyozhin94} and \citet{Junde11}, respectively.
Second, the gray and Monte Carlo gamma-ray transport schemes can lead to 
$\approx2$\% difference in the overall gamma-ray deposition rate. 
Difference in the amount of thermalized gamma-ray energy may further contribute to the
deviations of the light curves in the decay phase.

So far we have held the number of frequency bins at \stella's default of 
$N_\textrm{freq} = 40$.
In Figure \ref{fig:lc_Nfreq_comparison} we show the results of a convergence test
where we vary $N_\textrm{freq}$ from 4, 40, 400, to 15,000 in the \sedona\ models.
The bolometric light curves converge once $N_\textrm{freq} \ge 40$.
This experiment serves as an independent validation for the default choice of 
$N_\textrm{freq}$ in \stella, confirming the insensitivity of the light curve shape 
to the choice once the opacity variations are adequately resolved \citep{Paxton18}. 

Figure \ref{fig:explore_lcs} compares the bolometric light curves from the three
exploratory models.
For the II-P models on the left panel, we performed the \stella-to-\sedona\ handoff 
at day 30 instead of day 20 to allow more time for the inner ejecta to reach homology.
The M7.9\_R375\_E0.23\_Ni4.3 model represents an ejecta with an 
unusually low $E_{\rm Exp}$.
The low $E_\textrm{Exp}$ leads to a relatively slow expansion,
and a long plateau of $>120$\,days.
M16.5\_R533\_E4.6\_Ni13 has high $E_{\rm Exp}$ of $4.6\times 10^{51}$\,erg
and $M_\textrm{Ni}$ of 0.13\,$M_{\odot}$, giving rise to the overall much more luminous SN.
Even with extreme model parameters, the light curves from the two codes remain in agreement, 
with a comparable level of difference to the typical nickel-rich
models.

The right panel of Figure \ref{fig:explore_lcs} shows the light curves from the 
Type II-b-like model Stripped\_M4.7\_E1.0\_Ni3.
Unlike the II-P models, the ejecta quickly cooled and settled onto the radioactive
decay tail by day 50.
Between day 20$-$30, the ejecta is moderately optically thick with an integrated 
optical depth $\tau_\textrm{R} = \int \kappa_\textrm{R} \rho dr$\,$\approx$ 30 - 100,
where $\kappa_{\rm R}$ is the Rosseland mean opacity.
The release of the initial thermal radiation leads to a higher bolometric luminosity 
than the instantaneous nuclear decay luminosity.
The slopes of the decay tails are steeper than the analytical prediction
that assumes total trapping,
consistent with the expectation of gamma-ray leakage from the ejecta
\citep{Clocchiatti97,Wheeler15,MA20}.
The steeper slope from the \stella\ model is likely due to an underestimation 
of gamma-ray deposition by the \citet{SSH95} scheme. 
After day $\approx$80, the ejecta's total gamma-ray optical depth drops to order of unity.
The gamma-ray deposition rate therefore depends sensitively on the actual value of 
the gamma-ray opacity used.
In the other (II-P) models, gamma-ray optical depths 
are high even at late times. The near-total deposition of gamma-rays renders 
the deposition rate insensitive to the exact opacity value.
In addition, assuming a gray, purely absorptive gamma-ray opacity can lead to substantial differences in 
the gamma-ray deposition rate as compared to a frequency-dependent calculation, 
especially in the outer ejecta \citep[Figure 7]{Wilk19}.
Nevertheless, the light curves agree within $\approx$10\%.

\begin{figure}[t]
\includegraphics[width=\columnwidth]{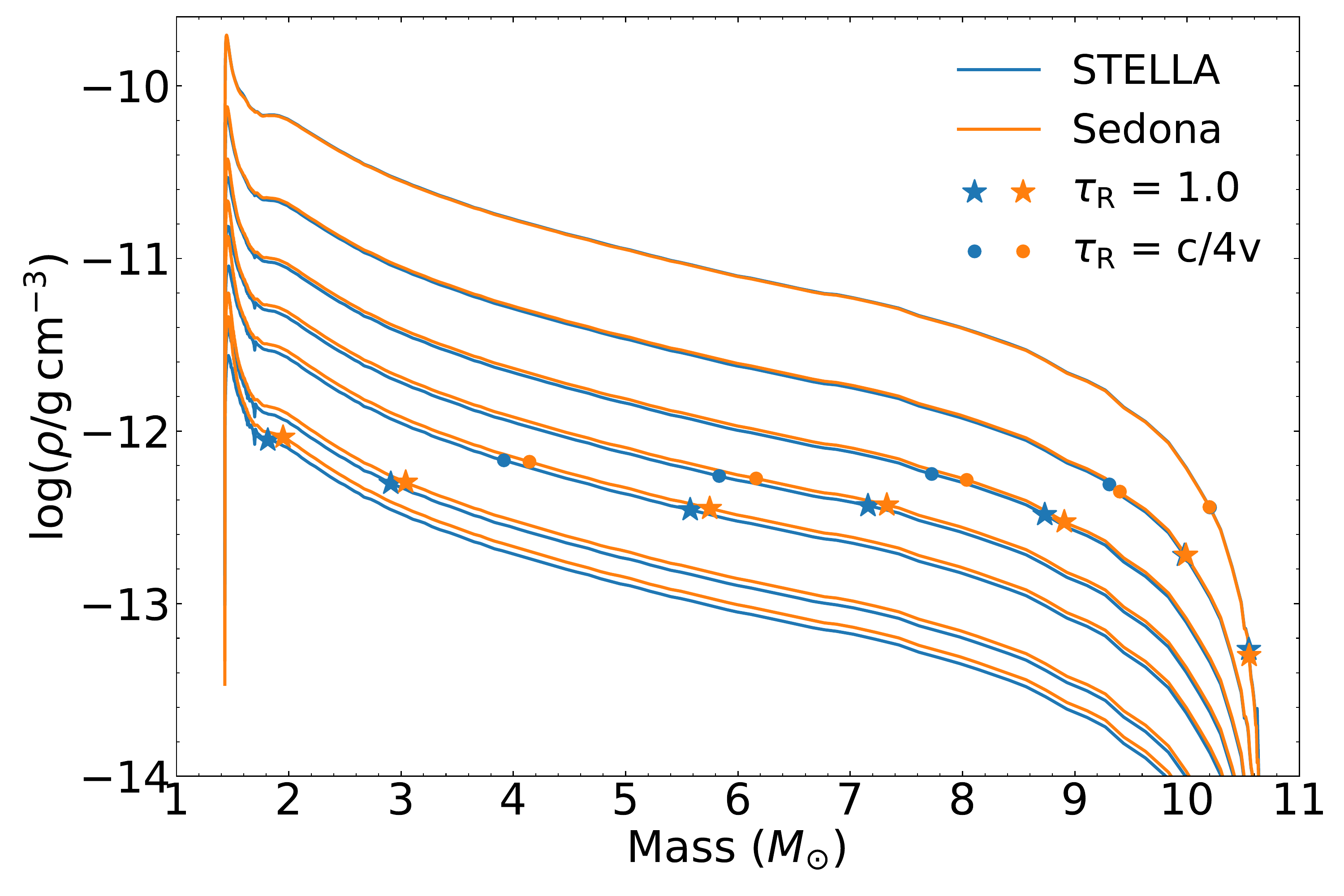}
\caption{Evolution of density in the M9.3\_R433\_E1.0\_Ni3.0 models.
         The curves correspond to, from top to bottom, profiles at
         day 20, 30, 40, 50, 60, 80, and 90.
         The stars and points mark the locations of the photosphere 
         and where $\tau_\textrm{R} = c/4v$, respectively.
         Assuming homology in \sedona\ only introduces $\approx$10\% 
         relative difference with the full hydrodynamical evolution in \stella.
         } \label{fig:dens_profile_comparison}
\end{figure}

\subsection{Ejecta Thermal Structures}
\label{sec:ej_structures}
We now compare the thermodynamical states of the ejecta calculated from the two different codes. 
Deep in the interior, electron scattering dominates the opacity and traps the radiation.
In the absence of radioactive heating, the gas temperature then evolves adiabatically 
as the ejecta expands.
We show the gas temperature evolution of the M9.3\_R433\_E1.0 model
in the left panel of Figure~\ref{fig:temp_profile_comparison}. 
The adiabatic evolution is shown as black dashed lines.
Adiabaticity breaks down when the radiative diffusion timescale is 
comparable to the dynamical time at $\tau \approx c/4v$. 
The mass coordinate locations $m_\textrm{c/4v}$ where 
$\tau_\textrm{R}(m_\textrm{c/4v}) = \int_{m=m_\textrm{c/4v}}^{M_\textrm{ej}} \kappa_{\rm R}(r) \rho(r)dr = c/4v$
are denoted by circular points.
Below these locations, the adiabatic scaling predicts the temperature evolution.
The locations of the photosphere $m_\textrm{ph}$, which are defined as 
$\tau_\textrm{R}(m_\textrm{ph}) = 1$, are marked by the star symbols.

\begin{figure}[t]
\includegraphics[width=\columnwidth]{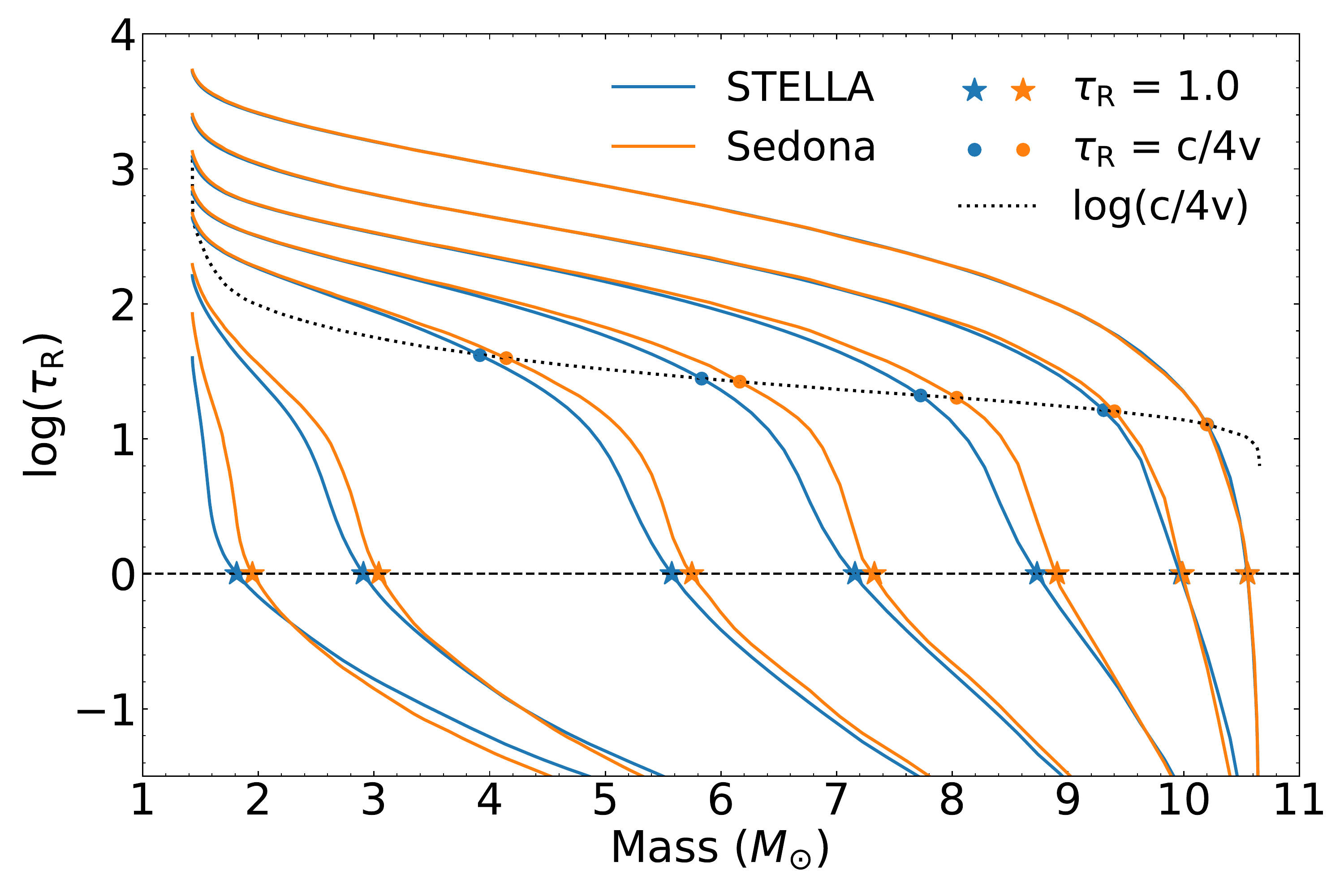}
\caption{Evolution of the integrated optical depth in the 
         M9.3\_R433\_E1.0\_Ni3.0 model.
         The curves correspond to, from top to bottom, profiles at
         day 20, 30, 40, 50, 60, 80, and 90.
         The stars and points mark the locations of the photosphere and 
         where $\tau_\textrm{R} = c/4v$, respectively.
         } \label{fig:tau_profile_comparison}
\end{figure}

We can also derive a temperature prediction assuming the radioactive energy release
(in forms of both gamma-rays and positrons) is locally deposited.
Mathematically, we adopted the nuclear energy production rate from Equation (13) of 
\citet{Nadyozhin94}. 
On the right panel of Figure \ref{fig:temp_profile_comparison},
such analytical scaling again reproduces the inner ejecta temperature 
where radiative trapping dominates ($\tau_\textrm{R} \ge c/4v$).
At $\tau_\textrm{R} < c/4v$ (above the circular points), we expect the temperature
structures to be modified by the transport of radiation.

To understand the slight offsets of the \stella\ and \sedona\ temperature profiles
near the photosphere,
we show in Figure \ref{fig:dens_profile_comparison} the density evolution of the
same M9.3\_R433\_E1.0\_Ni3.0 model.
While the assumption of homology in \sedona\ follows the density evolution to within
$\approx$10\% of \stella, the residual hydrodynamical acceleration in \stella\ 
after model handoff lowers the overall optical depth.
As a result, radiative cooling via diffusive losses is slightly more efficient and 
the recombination front and photosphere recede deeper into the ejecta.
The resultant offsets in the photospheric locations are also illustrated in Figure~\ref{fig:tau_profile_comparison}.

\begin{figure}
\includegraphics[width=\columnwidth]{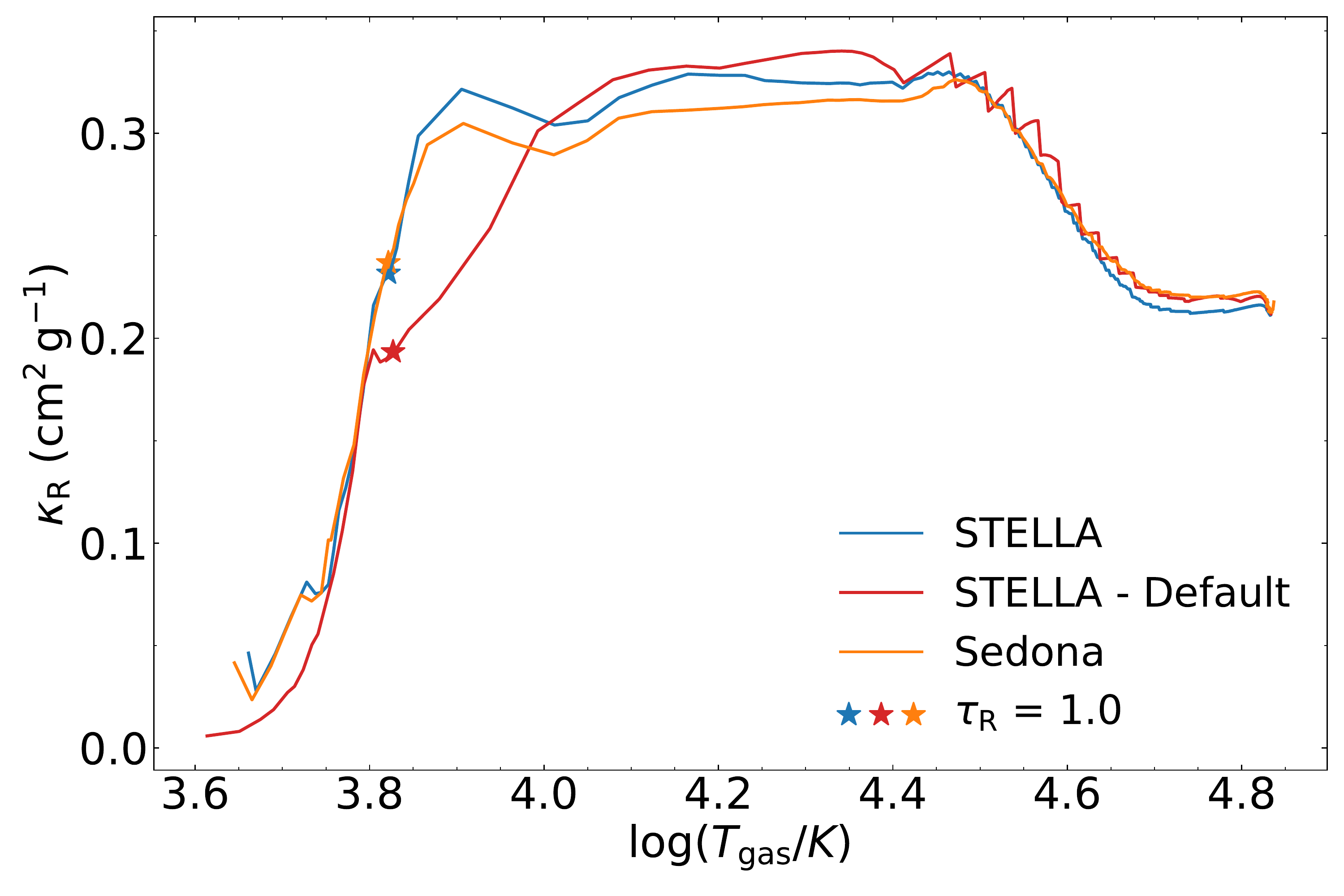}
\caption{Rosseland mean opacity as a function of gas temperature in the
        M9.3\_R433\_E1.0\_Ni3.0 model at the moment of model handoff (day 20).
        The star markers represent the locations of the photosphere.
        The sawtooth patterns at $\log(T_\textrm{gas}) \ge 4.4$ 
        are the result of the coarse zone resolution (N$_\textrm{reduced}$ = 50)
        of the opacity tables in the default version of \stella.  
        } \label{fig:kappa_temp_comparison}
\end{figure}

A comparison of opacity as a function of gas temperature at the time of handoff
is shown in Figure \ref{fig:kappa_temp_comparison} for the 
M9.3\_R433\_E1.0\_Ni3.0 model.
Even though the radiation transport schemes are very different between the codes, 
the opacity variations they encompass are mostly consistent.
The opacity from the default version of \stella\ packaged with \mesa\ 
is also shown --
the hydrogen recombination front is bracketed only by two temperature grid
points at $\log(T_{\rm gas}/K)$ of 3.8 and 4.0, causing large interpolation
errors in between.
Nevertheless, this offset in opacity does not introduce significant deviations
in the effective temperature,
resulting in overall very similar light curves.
At $\log(T/K) > 4.0$, the difference in opacity can be attributed to 
the different atomic line lists and analytical expressions for the bound-free
and free-free opacity.

\subsection{Towards Monte Carlo Radiation Hydrodynamics}

While the main focus of this work is to verify the reliability of different
radiation transport methods in light curve modeling,
the Monte Carlo approach of \sedona\ allows tallying of the radiative moments
at almost no additional costs.
A comparison of the radiative moments not only informs the robustness of both 
methods, it also serves as a numerical experiment leading towards 
multi-dimensional Monte Carlo radiation hydrodynamics. 

\begin{figure}
\includegraphics[width=\columnwidth]{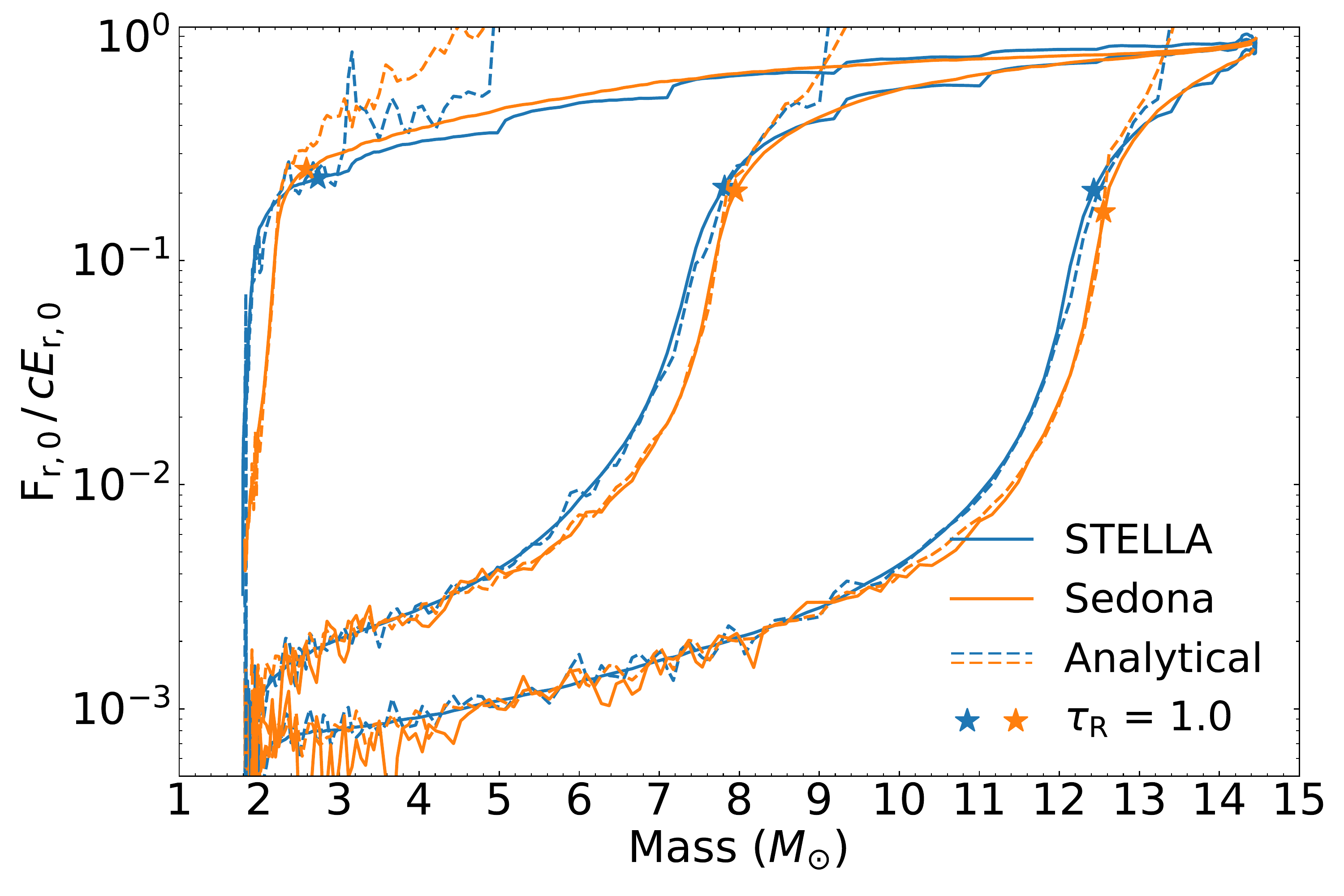}
\caption{Co-moving frame frequency-integrated flux factor at 
        day 120, 80, and 50 (top to bottom) of the 
        M12.7\_R719\_E0.84\_Ni4.8 model.
        Locations of the photosphere are marked with the star symbols.
        The analytical predictions based on Fick's law of radiative diffusion
        are plotted as dashed lines.
        } \label{fig:f2E_comparison}
\end{figure}

In the mixed-frame radiation transport formalism of \sedona, 
the radiation energy density $E_\textrm{r}$, radiative flux $F_\textrm{r}$, 
and radiation pressure $P_\textrm{r}$ are tracked in the lab frame.
In spherical symmetry, non-radial quantities vanish and only the radial components
of the radiative flux and pressure are followed.
The lab frame quantities are Lorentz-transformed into the co-moving frame 
quantities $E_\textrm{r,0}$, $F_\textrm{r,0}$, and $P_\textrm{r,0}$
\citep{MihalasMihalas99}.
The profiles of the co-moving frequency-integrated flux factor 
$F_\textrm{r,0}/cE_\textrm{r,0}$ are plotted in Figure~\ref{fig:f2E_comparison}
for the M12.7\_R719\_E0.84\_Ni4.8 model.

In the deep interior of the ejecta, radiation is in the diffusive regime and 
is nearly isotropic.
The net radial flux is therefore small compared to the energy density.
In this region, the flux factors computed by the two radiation transport methods 
agree well with each other,
although the MC scheme shows statistical noise.
The fluctuation in \sedona's flux factor originates from the numerical difficulty of
representing a near-zero value by the sum of numerous discrete values
close in magnitudes but opposite in signs.
We also include the analytical predictions based on the Fick's law of radiative
diffusion
$F_\textrm{r,0}/cE_\textrm{r,0} = -(1/3\kappa_\textrm{R}\rho) d\ln(E_\textrm{r,0})/dr$
as dashed lines, 
which accurately track the flux factor below the photosphere.
The non-monotonicity in the analytical predictions of the \stella\ model 
is the result of discretization errors in computing the $d\ln(E_\textrm{r,0})/dr$
term from the spatially varying zone sizes.

Near the photosphere, the factor quickly converges to a larger value
with a much lower level of MC noise.
Above the photosphere, radiation approximates free-streaming and the flux factor 
approaches unity.
In this optically thin region, Fick's law no longer holds and the analytically 
predicted flux factor diverges from the true value as expected.

In the diffusion-dominated regime, \citet{Roth15} have proposed a 
variance-reduction MC flux estimator based on the divergence of radiation pressure,
i.e., $ F_\textrm{r,0} = c \nabla \cdot P_\textrm{r,0}/\kappa_\textrm{R}\rho
\approx c \nabla \cdot E_\textrm{r,0}/3\kappa_\textrm{R}\rho$.
When radiation is isotropic, following the scalar quantity $E_\textrm{r,0}$
with discrete photon packets is less prone to statistical noise than
tallying the vector quantity $F_\textrm{r,0}$.
We confirm that this noise reduction estimator reproduces exactly the 
analytical predictions shown in Figure \ref{fig:f2E_comparison}. 

We observe that both the \stella\ and \sedona\ models agree well with 
its Fick's law predictions deep in the ejecta, and evolve smoothly outward.
It shows that the Monte Carlo radiation transport method holds promises in radiation
hydrodynamical applications, 
where the radiative moments can be used as source terms
coupled to gas dynamics evolution.

\section{Summary} \label{sec:summary_discussion}

In this work, we conduct the first detailed comparison of the light curves
and ejecta structures produced by the moment-based radiation hydrodynamical
code \stella\ and the particle-based, Monte Carlo radiation transport code \sedona.
Radiation transport methods based on radiative moments and discrete photon packets are
fundamentally different.
With ejecta models derived from realistic progenitor stellar evolution, 
this comparison serves as an important evaluation of two approaches to 
radiation transport modeling of supernovae. 

During the plateau phase, the light curves generated by the
two codes agree to within $\approx$5\%. 
In the late-time radioactive decay phase, the difference is at $\approx$3\% level,
owing to the different handling choices for the deposition of positron kinetic energy 
from the cobalt decays.
The agreements provide confidence in the fidelity of both codes and radiation 
transport approaches.
\sedona's homology assumption in evolving the ejecta structures is also found to 
reproduce the ejecta's full thermodynamical evolution fairly well. 
The residual hydrodynamical acceleration in \stella\ only induces $\approx$10\% differences
in the density profiles, and resultant slight offsets in the photospheric locations.

Although Monte Carlo radiation transport is highly adaptable and is not subject to any
specific closure relations, a large number of photon packets is required to
suppress the statistical noise below an acceptable level. For instance, 
a typical \sedona\ run including radioactive decays follows the histories of $\sim$10$^{8}$ 
photon packets. 
The high computational demands limit the scope of parameter surveys that are 
tractable with \sedona.
Current progress is being made to break the efficiency barrier by optimizing the 
radiation transport in the diffusive regime within \sedona.

The flexibility of the Monte Carlo approach to radiation in \sedona\ also tracks the
radiative moments (energy density, flux, and pressure) and verifies the quality of the 1D closure scheme in \stella.
We find that \stella's moment method tracks the transition from the minuscule flux 
in the diffusive regime to the free-streaming flux in the optically thin regime
with comparable accuracy to the Monte Carlo approach. 
The present work demonstrates that it is promising to extend the Monte Carlo framework towards full radiation hydrodynamical simulations.

\acknowledgments

We are grateful to the anonymous referee for the constructive comments that 
improved the content of this paper.
We thank Bill Paxton for the valuable inputs and discussions during the course of the study.
This research project has also benefited from interactions with
Evan Bauer, Siva Darbha, David Khatami, Hannah Klion, Thomas Kupfer, 
Andrew MacFadyen, Maryam Modjaz, Abigail Polin, Josiah Schwab, and Christopher White.
This research was funded by the Gordon and Betty Moore Foundation through Grant GBMF5076.
This research was supported in part by the National Science Foundation under Grant No. NSF PHY-1748958. J.A.G. is supported by the NSF GRFP under grant number 1650114.
We acknowledge support from the Center for Scientific Computing from the CNSI, 
MRL: an NSF MRSEC (DMR-1720256) and NSF CNS-1725797.

%

\vspace{5mm}


\software{
          Jupyter \citep{Kluyver16},
          Matplotlib \citep{Hunter07},
          \mesa\ \citep{Paxton2011,Paxton2013,Paxton2015,Paxton18,Paxton19},
          NumPy \citep{Oliphant06},
          SciPy \citep{Jones01},
          \sedona\ \citep{Kasen06}, \stella\ \citep{Blinnikov98,BS04,BBP05,Blinnikov06}
          }

\bibliographystyle{aasjournal}
\bibliography{biblio}



\end{document}